\documentclass[%
 aip,
 jmp,%
 amsmath,amssymb,
preprint,%
]{revtex4-1}

\usepackage{graphicx}
\usepackage{dcolumn}
\usepackage{bm}

\begin{document}

\preprint{AIP/123-QED}

\title[]{Capacitive displacement method to determine the longitudinal piezoelectric coefficients of single crystals, ceramics and thin films}

\author{Desheng Fu$^{1,2,3}$\\
$^1$Department of Electronics and Materials Science, Faculty of
Engineering, Shizuoka University, 3-5-1 Johoku, Naka-ku, Hamamatsu
432-8561, Japan.\\
$^2$Department of Engineering, Graduate School of Integrated Science
and Technology, Shizuoka University, 3-5-1 Johoku, Hamamatsu
432-8561,
Japan.\\
$^3$Department of Optoelectronics and Nanostructure Science,
Graduate School of Science and Technology, 3-5-1 Johoku, Naka-ku,
Hamamatsu 432-8011, Japan.}


\email{fu.tokusho@shizuoka.ac.jp}
\homepage{http://www.grl.shizuoka.ac.jp/~ddsfu/}



\date{\today}

\begin{abstract}
Recent developments in piezoelectric films have heightened the need
for the reliable methods to correctly characterize their
piezoelectric coefficients. Here, we demonstrate that capacitive
displacement method can be used to determine  longitudinal
piezoelectric coefficients in both  bulk and thin film of
piezoelectric materials reliably and simply. Our technique allows
accurate  detection of elastic displacement at a level of 2 pm. This
achievement is of great interest and significance for the
development of piezoelectric integration technology in modern smart
devices.

\end{abstract}

\maketitle

Piezoelectrics convert  mechanical energy into electric energy or
vise versa, and show direct or converse
effects.\cite{Curie,Jaffe,Lines,Uchino} In the direct effect,
electric charge is produced at the surface of  piezoelectric crystal
in proportion to the applied stress. Conversely, as a suitably
oriented electric field is applied, the crystal changes shape
(strains) in proportion to the electric field. Such properties
provide piezooelectrics a wide spectra of applications such as
medical ultrasound imaging, fuel injecting in cars, ultra-precise
positioning in optics or mechanics and  sonar sensing in fish or
submarine detection, and reserach on piezoelectric is still a hot
topic of current material
sciences.\cite{FuAPL2007,FuPRL2008,FuAPL2008,Rodel,TazakiJPC2009,FuJPC2011,FuAPL2011,FuJPC2013}
With the combination of modern thin film integration technology and
nano-technology, piezolectrics can be used in nano-sensors,
-actuators or -transducers in modern smart
devices.\cite{Muralt1995,Muralt1997,Damjanovic,DeVoe} For these
applications, understanding the piezoelectric properties of
integrated thin film is of crucial importance.

Conventionally, to determine the piezoelectric coefficient of bulk
materials, resonant anti-resonant dynamic method, $d_{33}$-meter
method based on the static direct effect, or interferometry method
based on the static converse effect have been developed.\cite{Jaffe}
In principle, all these techniques can be used for the measurements
on thin
films.\cite{Pan,Lefki,Kanno,Kholkin,Christman,Xu,Durkan,Kuffer,FuJJAP,Park}
Practically, one can find many difficulties in applying these
techniques to measure the piezoelectric coefficient of thin films.
In the resonant anti-resonant dynamic method, when the film
thickness is on the order of $\sim$ 100 nm, the resonant and
anti-resonant frequencies  reach  approximately 10 GHz, which makes
pratical detection in a continuous frequency range extremely
difficult. In addition to the dimension limitation, there is also an
issue of substrate-constraint to be solved in the application of the
resonant and anti-resonant method in thin films. Currently, to
determine the piezoelectric coefficient of thin films, one usually
probes the elastic deformation of the film with an electric field by
the atomic force microscopy (AFM)\cite{Christman} or
interferometry.\cite{Pan,Kholkin} These two appoaches can provide a
resolution of $\sim$ pm, however, AFM-tip-field electrostatic
interactions \cite{Christman} or  the bending effects of substrate
\cite{Kholkin}  may be a barrier for the  reliable measurements of
exact elastic deformation of the piezoelectric film. Therefore,
reliable determination of  the piezoelectric coefficient of thin
films is still a challenging issue at present.

Here, we demonstrate that  capacitive displacement method can be
used  to determine the longitudinal piezoelectric coefficients in
both thin films  and  bulk materials reliably and simply. Our
approach can correctly detect  displacement as small as 2 pm and
offers an opportunity of improved measurements of piezoelectric
coefficient of thin films.

\begin{figure}
\includegraphics[width=12cm]{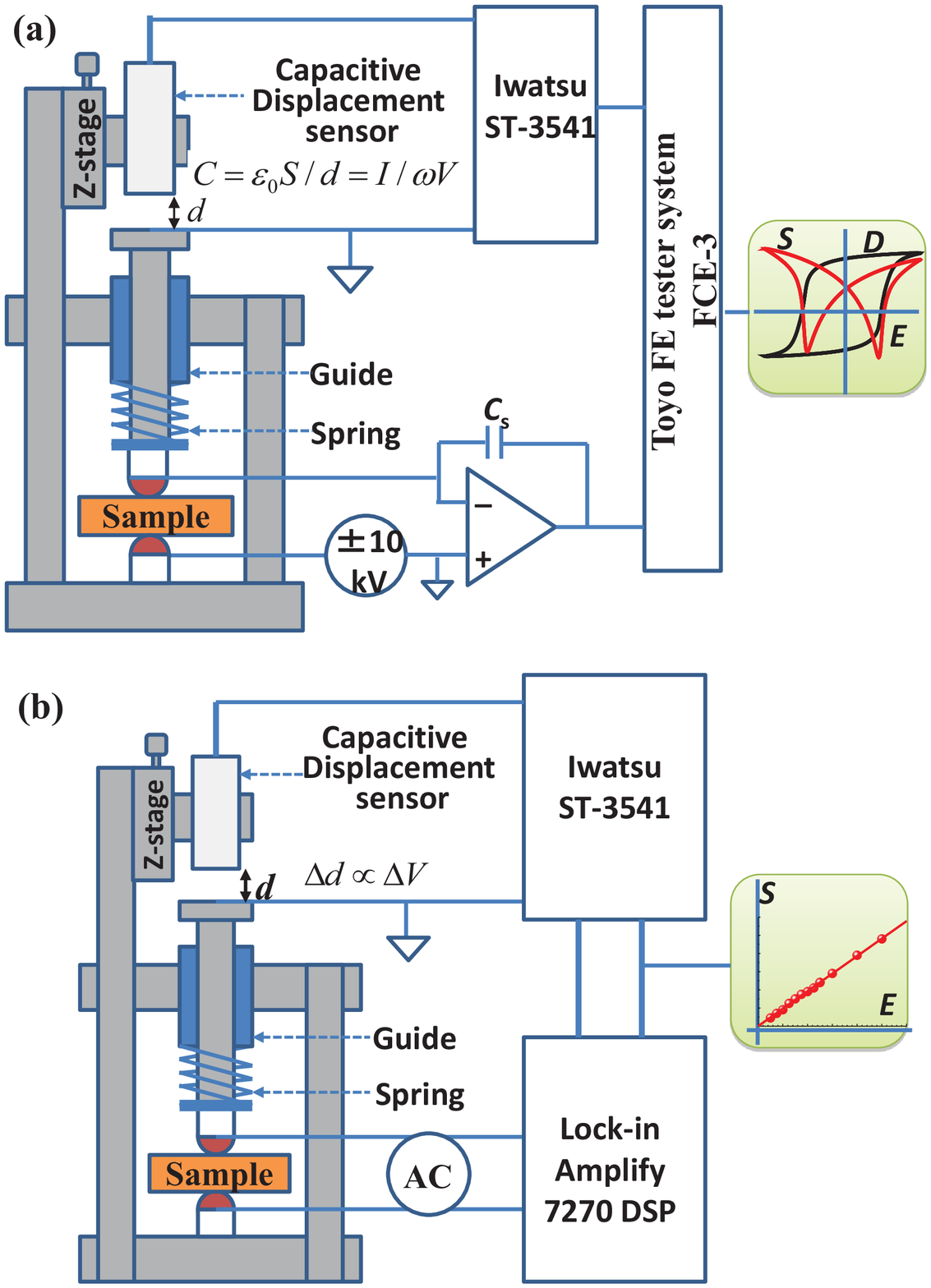}
\caption{\label{Fig1} Setup of capacitive displacement method  to
measure to the displacement of piezoelectric induced by an electric
field: (a) Quasi-static method that allows simultaneously to measure
the displacement and dielectric polarization as a function of
bipolar or unipolar electric field and provides a resolution of
$\sim$ nm for the displacement measurement. (b) AC method that
employs  the lock-in amplifier and has a higher resolution ($\sim$
pm.) }
\end{figure}

Figure \ref{Fig1} shows the schematic of our capacitive displacement
method, in which the displacement is detected through detecting the
change in air capacitance $C$ forming between the capacitive
displacement sensor and the object to be measured. Here, we have
\begin{equation}\label{eq1}
    C=\varepsilon_0 S/d,
\end{equation}
where $\varepsilon_0$ is permittivity of vacuum, $S$ is the area of
senor, $d$ is the distance between the sensor and the object to be
measured, respectively. When voltage ($V$) with frequency $f$ is
applied between the sensor and the object to be measured,
\begin{equation}\label{eq2}
    V=I/(2\pi f C)=d I /(2\pi f \varepsilon_0 S).
\end{equation}

As can be seen from equation (\ref{eq2}), when the current $I$
flowing through the sensor is maintained at a constant value, $d$ is
proportional to the output voltage $V$. Therefore, one can measure
the displacement by measuring the output value of the capacitive
sensor, which in our setup is performed by a commercial displacement
meter Iwatsu 3541,  providing a probing range of $\pm 25 \mu$m with
a resolution of approximately 1 nm. As shown in Fig. \ref{Fig1}, the
sample is held between two cylindrical electrodes that provide point
contact between the sample and the cylindrical electrode, allowing
free motion of the sample. When the electric field is applied to the
piezoelectric sample, its longitudinal deformation is transfered to
the bottom electrode of the air capacitor by a spring and a linear
bush. Linear bush also provides a solid guide for the system.  Using
a Z-stage, one can adjust the air capacitance to set the output of
the displacement meter to zero as the sample is in a resting state
without an elastic deformation, which allows one to directly read
the exact displacement using the displacement meter.

\begin{figure}
\includegraphics[width=12cm]{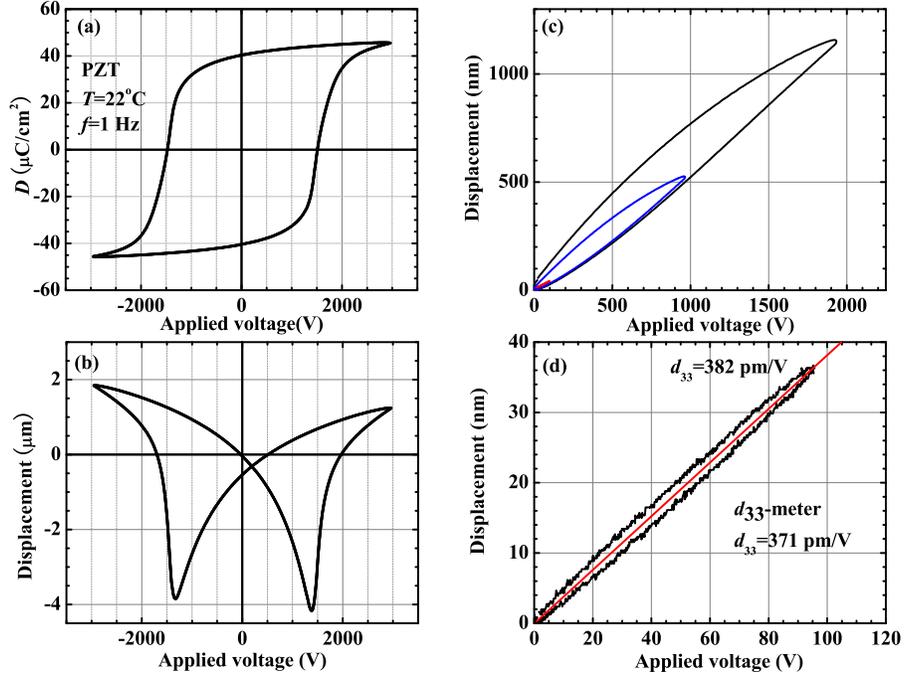}
\caption{\label{Fig2} (a) D-E hysteresis loop of the industrial PZT
ceramics with a thickness of 1 mm obtained by a bipolar electric
filed. (b) Displacement loop obtained by the quasi-static method
simultaneously with the D-E loop. (c) and (d) show the displacement
obtained by the quasi-static method for the large and small unipolar
electric fields. $d_{33}$ value was estimated by the slope of red
solid line shown in figure (d) and its value is also indicated. For
comparison, $d_{33}$ value measured by the $d_{33}$-meter is also
shown in the  figure. }
\end{figure}

In the setup shown in Fig.\ref{Fig1}(a), the displacement meter is
combined with a commercial ferroelectric tester (Toyo FE tester
system FCE-3), which allows one to measure the D-E hysteresis loop
and the strain loop simultaneously for a bipolar electric field.
Reliable running of the system is  confirmed by the results shown in
Fig. \ref{Fig2}, in which commercial Pb(Zr$_{1-x}$Ti$_x$)TiO$_3$
(PZT) ceramics with a composition near the morphotropic phase
boundary (MPB) was used. This ceramics sample shows a typical
butterfly shape of strain loop for a bipolar field. For a unipolar
field, it shows a large displacement of 1150 nm for an applied
voltage of 1900 V. To examine whether the measured displacement is
correct or not, we compare the result obtained at a lower voltage
with that measured by a $d_{33}$-meter ( Model ZJ-6B Quasi-Static
Piezo $d_{33}$/$d_{31}$-meter made by Institute of Acoustics,
Chinese Academy of Sciences.). For the applied voltage of 95.7 V, we
have detected a displacement of 36.5 nm for the PZT ceramics sample.
The ratio of displacement divided by the applied voltage is 382
pm/V, which provides an rough estimate of piezoelectric coefficient
$d_{33}$ of the sample. This value is in good agreement with the
value of 371 pm/V obtained by a $d_{33}$-meter when considering that
domain motion still has a small contribution to the total strain at
this small applied voltage, which is evident from the small
hysteresis of the strain curve. From these results, we can conclude
that our capacitive displacement measurement system is highly
reliable.

\begin{figure}
\includegraphics[width=12cm]{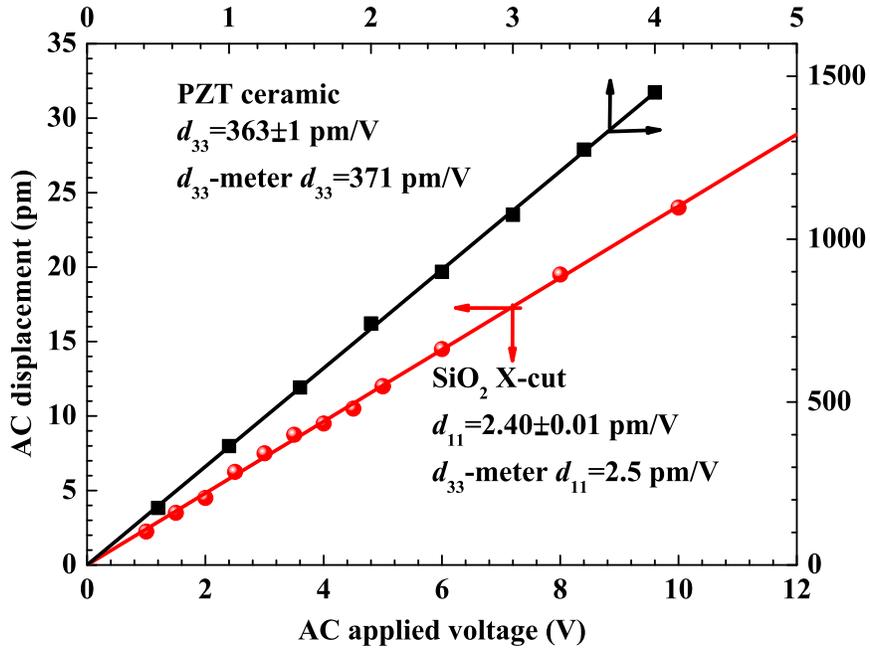}
\caption{\label{Fig3} Displacement obtained by the AC capacitive
method as a function of the AC electric field for  the PZT ceramics
sample used in Fig.\ref{Fig2} and X-cut quartz single crystal. The
longitudinal piezoelectric coefficients were obtained by a simple
linear fit with parameters indicated in the figure. For comparison,
values measured by the  $d_{33}$-meter  are also shown. }
\end{figure}

From Fig.\ref{Fig2}(d), one can see that the noise level of the
method shown in Fig.\ref{Fig1}(a) is on the order of approximately 1
nm. Using this method it is not easy to measure the displacement of
thin film correctly, since the electric field induced displacement
in the thin film normally falls within the range of noise level. To
greatly improve the resolution of displacement measurement, we have
proposed an AC method that is combined with a lock-in technique. The
setup implementing the AC method is shown schematically in
Fig.\ref{Fig1}(b). In this method, an AC field lower than the
coercive field of the ferroelectric material is applied to the poled
sample. This AC field leads to an AC displacement of the
piezoelectric sample with the same frequency. The output of the AC
displacement meter is then fed into a lock-in amplify (SIGNAL
RECOVERY model 7265) to reduce the noise. Using this approach, we
can detect an AC displacement of $\sim$ pm reliably. To examine
reliability of this method, we have used X-cut quartz as a stand
because X-cut quartz is a piezoelectric crystal but not a
ferroelectric crystal, its piezoelectric coefficient is not affected
by polarization domain and has been widely reported in the
literatures.\cite{Bottom,Pan,Kholkin} The AC displacement of X-cut
quartz as a function of AC applied voltage is shown in
Fig.\ref{Fig3}. From the linear fit, $d_{11}$ of X-cut quartz is
evaluated to be $2.40\pm0.01$ pm/V. While the $d_{33}$-meter gives a
value of 2.5 pm/V for the same sample. The known values for X-cut
quartz are 2.3-2.4 pm/V.\cite{Bottom,Pan,Kholkin} Comparing these
values we can conclude that our AC displacement measurement system
is highly reliable. The reliability of the measurement system is
further confirmed by the result of the PZT ceramics sample used in
Fig.\ref{Fig2}, which is also shown in Fig.\ref{Fig3}. The AC method
gives a $d_{33}$ value of $362\pm 1$ pm/V for the applied AC voltage
up to 4 V. This  $d_{33}$ value is consistent with that shown in
Fig. \ref{Fig2}(d) and obtained by the $d_{33}$-meter. Also, it can
be seen clearly from Fig.\ref{Fig3} that at least 2 pm displacement
can be exactly determined by our AC capacitive displacement
measurement.

\begin{figure}
\includegraphics[width=12cm]{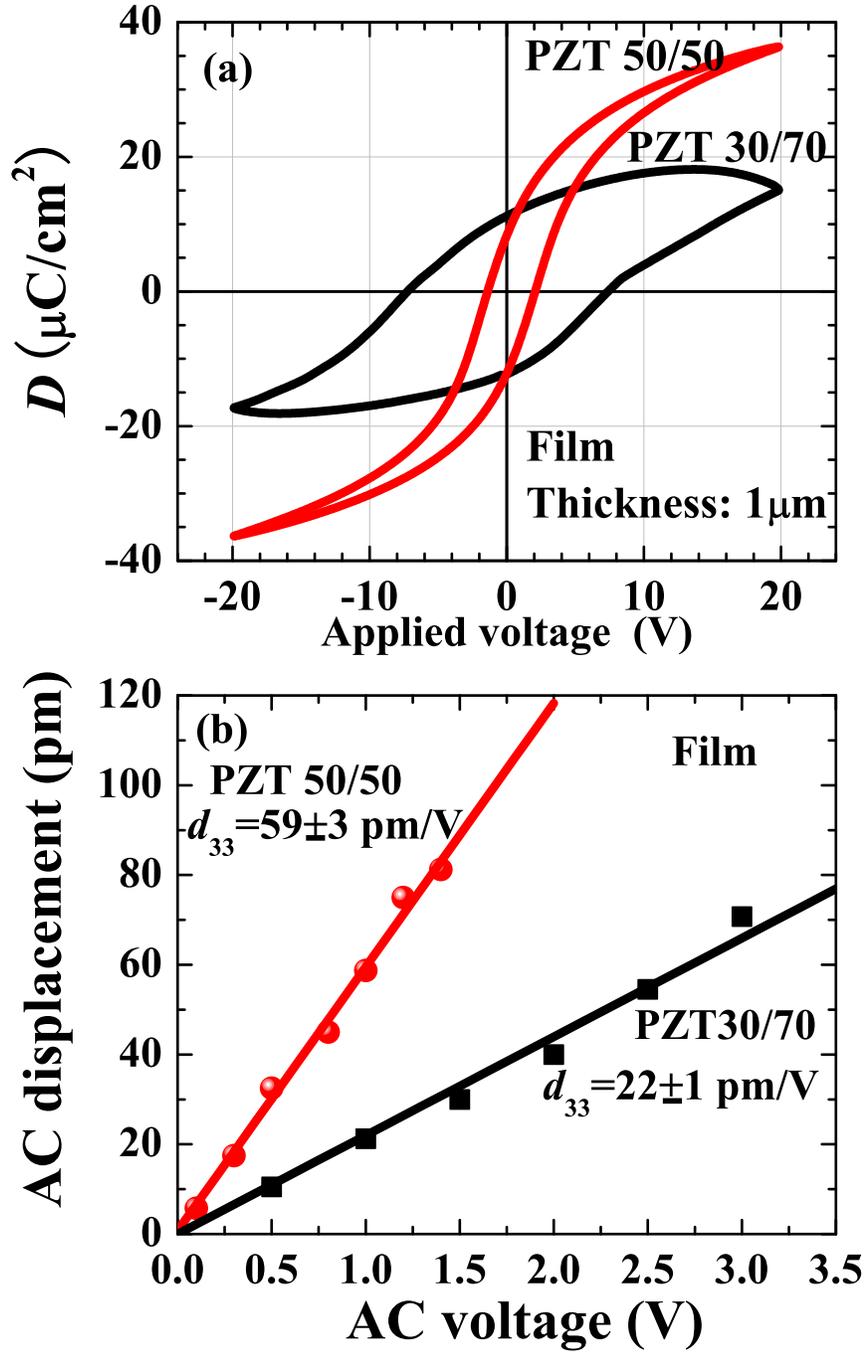}
\caption{\label{Fig4} (a) D-E  hysteresis loops of PZT thin
films.(b) AC displacement obtained by the AC capacitive method as a
function of the AC electric field for PZT thin films. The
longitudinal piezoelectric coefficients are also indicated in the
figure. }
\end{figure}

After examining the reliability of our AC capacitive displacement
method, we then used it to test two kinds of PZT thin films
(PZT50/50 and PZT30/70). PZT50/50 and PZT30/70 films have a
composition of $x=0.5$ and $x=0.7$, respectively. These PZT films
were prepared by a chemical solution deposition on a silicon
substrate with a Pt bottom electrode for capacitor fabrication. To
form a capacitor, a top electrode with diameter of 1 mm was coated
by  Au. For the convenience of measurements, the back side of
silicon substrate without PZT film was also coated by Au and
connected to the Pt bottom electrode of PZT film. Figure
\ref{Fig4}(a) shows the D-E hysteresis loop of these two films, both
of which have a similar value of remanent polarization of
approximately of 10 $\mu$C/cm$^2$, indicating that these films are
suitable for the piezoelectric measurements. The coercive fields
$E_c$ were measured to be 2 V/$\mu$m and 7.3 V/$\mu$m for the
PZT50/50 and PZT30/70 films, respectively. We then used an AC field
less than $E_c$ to measure the AC displacement. The measurements
were carried after D-E loop measurements which provided poling for
the PZT films. The results are shown in Fig.\ref{Fig4}(b). $d_{33}$
values of PZT50/50 and PZT30/70 films were estimated to be $59\pm 3$
pm/V and $22\pm 1$ pm/V, respectively. These results suggests that
PZT film with composition near MPB exhibits a large piezoelectric
response that has Ti-rich composition. The lower $d_{33}$ values of
PZT thin film as compared with ceramics samples may be affected by
many factors such as film quality, orientation, poling process.
However, detailed investigation of  these effects is beyond the
scope of the present study.

In summary, we have developed  capacitive displacement measurement
method to determine the longitudinal piezoelectric coefficient in
both bulk and thin films reliably and simply. Using the AC method
 combined with the lock-in technique, we can precisely determine
displacement as sall as  2 pm. This approach is expected to provide
new insight into the piezoelectric properties of thin films which is
still a challenging issue for the development of piezoelectric
integrated technology.


\begin{acknowledgments}
I thank Professor H. Suzuki in Shizuoka University for his kindly
providing the PZT thin films for the test of my system.
\end{acknowledgments}


%

\clearpage

\clearpage

\clearpage

\clearpage



\end{document}